\RequirePackage{fix-cm}
\documentclass[smallextended]{svjour3}       
\smartqed  
\usepackage{graphicx}
\usepackage{a4wide}
\usepackage{epsfig}
\usepackage{epsf}
\usepackage{psfrag}
\usepackage[affil-it]{authblk}
\usepackage{latexsym}
\usepackage{amsfonts}
\usepackage{amsmath}
\usepackage{mathrsfs}
\usepackage{amssymb}
\usepackage[activeacute,english]{babel}
\usepackage[noadjust]{cite}

\begin{document}

\title{Gravitational entropy of Kerr black holes
}

\titlerunning{Gravitational entropy of Kerr black holes}        

\author{Daniela P\'erez         \and
        Gustavo E. Romero 
}


\institute{Daniela P\'erez and Gustavo E. Romero \at
              Instituto Argentino de Radioastronom{\'\i}a, C.C.5, (1984)\\
              Villa Elisa, Bs. As., Argentina \\
              Tel.: +54-221-482-4903\\
              Fax: +54-221-425-4909\\
              \email{danielaperez@iar-conicet.gov.ar}           
           \and
           Gustavo E. Romero \at
              Facultad de Ciencias Astron\'omicas y Geof{\'\i}sicas, UNLP, Paseo del Bosque s/n\\
							CP (1900), La Plata, Bs. As., Argentina\\
							\email{romero@iar-conicet.gov.ar}}

\date{Received: date / Accepted: date}

\maketitle

\begin{abstract}
Classical invariants of General Relativity can be used to approximate the entropy of the gravitational field. In this work, we study two proposed estimators based on  scalars constructed out from the Weyl tensor, in Kerr spacetime. In order to evaluate Clifton, Ellis and Tavakol's proposal, we calculate the gravitational energy density, gravitational temperature, and gravitational entropy of the Kerr spacetime. We find that in the frame we consider, Clifton et al.'s estimator does not reproduce the Bekenstein-Hawking entropy of a Kerr black hole. The results are compared with previous estimates obtained by the authors using the Rudjord-Gr$\varnothing$n-Hervik approach. We conclude that the latter represents  better the expected behaviour of the gravitational entropy of black holes. 
\keywords{Gravitation \and General Relativity \and Black Holes \and Entropy}
\end{abstract}

\section{Introduction}
\label{intro}

The study of quantum field theory in the external gravitational field of a black hole led to the proposal that black holes have temperature and entropy. The radiation emitted by the black hole is thermal and it has a temperature given by \cite{Hawking1} \cite{Hawking2}: 
\begin{equation}
\Theta_{\rm H} = \frac{\hbar \ c^{3}}{8  \pi \: G \: k_{\rm B}\: M},
\end{equation}
where $k_{\rm{B}}$ is Boltzmann's constant, $G$ is the constant of gravitation, $c$ stands for the speed of light, and $\hbar=h /2 \pi$. As usual, $h$ is the Planck's constant.

Bekenstein \cite{Bekenstein1} \cite{Bekenstein2} noticed that the properties of the area of the event horizon of a black hole resemble those of entropy and proposed the following relation:
\begin{equation}
S_{\rm BH} = \frac{A}{4  \ l_{\rm P}^{2}}.
\end{equation}
Here, $S_{\rm BH}$ is the entropy of the black hole, $A$ is the area of the event horizon, and $l_{\rm{Pl}}^{2}= G\hbar c^{-3}$ is the Planck area. A generalized second law of black hole thermodynamics was also derived by Bekenstein \cite{Bekenstein3}. Bardeen, Carter, and Hawking \cite{Bardeen} formulated the four laws of black hole physics, which are similar to the four laws of thermodynamics.

Since black holes can be fully described in terms of the gravitational field, it seems reasonable to associate an entropy with the gravitational field itself. In absence of a theory of quantum gravity, a statistical measure of the gravitational entropy is not possible. Instead, approximations might be obtained using classical invariants of General Relativity. In a pioneer work, Penrose \cite{Penrose} proposed the so-called Weyl Curvature Conjecture: scalars constructed out from the Weyl tensor might be used to quantify the entropy of the gravitational field.

Several authors have tried to implement Penrose's proposal. For instance, Rudjord, Gr$\varnothing$n and Hervik \cite{Gron} defined a classical estimator for the gravitational entropy based on the ratio of the Weyl and Kretschmann scalars. They demanded that their estimator integrated over the event horizon of a black hole be equal to the Bekenstein-Hawking entropy \cite{Bekenstein3}.

Recently, Clifton, Ellis and Tavakol \cite{Clifton} offered a novel definition for the entropy of the gravitational field based on integrals over quantities constructed from the pure Weyl form of the Bel-Robinson tensor \cite{Bel1,Bel2,Robinson1,Robinson2}. In particular, they calculated the gravitational entropy for a Schwarzschild black hole, for a spatially flat Robertson-Walker geometry with scalar perturbations, and for the inhomogeneous Lema\^{i}tre-Tolman-Bondi solution.

The main goal of the present work is to calculate the gravitational entropy of a Kerr black hole using Clifton et al.'s proposal, to compare the results obtained with those corresponding for Rudjord et al's estimator, and to conclude which of them better approximates the entropy of the gravitational field in a rather complex spacetime such as Kerr's.

In the following section we provide a brief review of the work done by Rudjord et al. \cite{Gron} and Clifton et al. \cite{Clifton}. Throughout this paper we use geometrized units $G = c = 1$.

\section{Proposals for classical estimators of the gravitational field entropy}

\subsection{Weyl-Kretschmann estimator}
\label{sub:WK}

Rudjord and co-workers suggested an estimator for the gravitational entropy that naturally incorporates the Bekenstein-Hawking entropy. The entropy of a black hole can be described by the surface integral
\begin{equation}\label{Sint}
S_{\rm{\sigma}} = k \int_\sigma \vec \Psi \cdot \vec{d\sigma},
\end{equation}
where $\sigma$ is the surface of the horizon of the black hole and the vector field $\vec{\Psi}$ is
\begin{equation}
\vec \Psi = P \vec e_{{\rm{r}}},
\end{equation}
with $\vec e_{\rm{r}}$ a unitary radial vector. The scalar $P$ is defined in terms of the Weyl scalar ($W$) and the Kretschmann scalar ($R$):
\begin{equation}\label{p1}
P^{2} = \frac{W}{R} = \frac{C^{\alpha\beta\gamma\delta}C_{\alpha\beta\gamma\delta}}{R^{\alpha\beta\gamma\delta}R_{\alpha\beta\gamma\delta}}\;.
\end{equation}
To find an acceptable description of the entropy of a black hole, it is required that Eq. (\ref{Sint}) be equal at the horizon to the Bekenstein-Hawking entropy:
\begin{equation}
S_{\rm{\sigma}}=S_{\rm{BH}}.
\end{equation} 
The latter equation allows to calculate the constant $k$:
\begin{equation}
k=\frac{k_{\rm{B}}}{4l_{\rm{P}}^{2}}=\frac{k_{\rm{B}}c^{3}}{4G\hbar}.
\end{equation}

The entropy density can be then determined by means of Gauss's divergence theorem, rewriting Eq. (\ref{Sint}) as a volume integral:
\begin{equation}\label{den}
s = \mid{ \nabla \cdot \vec \Psi}\mid.
\end{equation}
Here, the absolute value brackets were added to avoid negative or complex values of entropy.

Rudjord et al. calculated the entropy density in Schwarzschild, de Sitter, and Schwarzschild-de-Sitter (SdS) spacetimes. The gravitational entropy in de Sitter spacetime is zero (the Weyl tensor vanishes), thus suggesting that the entropy of the cosmological horizon should be of non gravitational origin. Conversely, the Schwarzschild spacetime seems to have only gravitational entropy. Then, the Schwarzschild and de Sitter spacetimes were interpreted as two extreme cases of the SdS spacetime. This result seems to be a reasonable description of the gravitational entropy: large thermodynamical entropy in the early universe and large gravitational entropy around black holes.

\subsection{Bel-Robinson estimator}

Clifton, Ellis and Tavakol \cite{Clifton} defined the entropy of the gravitational field in analogy with the fundamental law of thermodynamics:
\begin{equation}
T_{\rm grav} dS_{\rm grav} = dU_{\rm grav} + p_{\rm grav}dV.
\end{equation}
Here, $T_{\rm grav}$, $S_{\rm grav}$, $U_{\rm grav}$ and $p_{\rm grav}$ stand for the effective temperature, entropy, internal energy, and isotropic pressure of the free gravitational field respectively, whereas $V$ is the spatial volume. Expressions for the effective energy density $\rho_{\rm grav}$ and pressure $p_{\rm grav}$ are derived from the Bel-Robinson tensor. The latter behaves like an effective energy-momentum tensor for free gravitational fields\footnote{Since the Bel-Robinson tensor has dimensions of $L^{-4}$, it is necessary to take the square of this quantity to obtain energy densities and pressures with the correct dimensionality \cite{Bonilla}.} \cite{Breton} \cite{Krish}. Depending on whether the gravitational field is Coulomb-like (Petrov type D spacetimes) or wave-like (Petrov type N spacetimes) different expressions are obtained for $\rho_{\rm grav}$ and $p_{\rm grav}$. For Coulomb-like gravitational fields, such as black hole spacetimes, the effective energy density and pressure in the free gravitational field take the form
\begin{eqnarray}
8 \pi \rho_{\rm grav} & = & 2 \alpha \sqrt{\frac{2{\cal W}}{3}}, \label{dens-entropy}\\
p_{\rm grav} & = & 0,
\end{eqnarray}
where $\alpha$ is a constant and ${\cal W}$ is the ``super-energy density'':
\begin{equation}
{\cal W} = \frac{1}{4}\left({E_{a}}^{b} {E^{a}}_{b}+{H_{a}}^{b} {H^{a}}_{b}\right),
\end{equation}
and $E_{ab}$ and $H_{ab}$ denote the electric and magnetic part of the Weyl tensor, respectively.

The temperature of the gravitational field is defined as a local quantity that reproduces the Hawking \cite{Hawking1}, Unruh \cite{Unruh}, and de Sitter temperatures \cite{deSitter} in the appropriate limits \cite{Clifton}. It has the following expression:
\begin{equation}\label{T-grav}
T_{\rm grav} = \frac{\left|\dot{u}_{a}z^{a}+H+\sigma_{ab} z^{a}z^{b}\right|}{2\pi}.
\end{equation}
Here $u_{a}$ is a timelike unit vector, $z^{a}$ is a spacelike unit vector aligned with the Weyl principal tetrad, $H = \Theta/3$ being $\Theta = \nabla_{a}u^{a}$ the expansion scalar and $\sigma_{ab} = \nabla_{(a}{u_{b}}_{)} + a_{(a}{u_{b}}_{)} - 1/3 \:\Theta h_{ab}$ is the shear tensor; $h_{ab}$ is the projection tensor $h_{ab} = g_{ab} -\left(u_{c}u^{c}\right)u_{a}u_{b}$.

Clifton and coworkers calculated the gravitational entropy of a Schwarzschild black hole. As expected, the estimator reproduces over the event horizon the Bekenstein-Hawking entropy. If this result can be generalized to other black hole spacetimes the estimator would be theoretically more attractive than the Weyl-Kretschmann one, since it is computed from basic laws and no phenomenological constants such as $k$ are involved

In the next section we present our calculations of the Bel-Robinson gravitational entropy for Kerr black holes.


\section{Bel-Robinson entropy estimator for Kerr black holes}

We consider a new form of the Kerr solution obtained by Doran \cite{Doran}. The line element in oblate spheroidal coordinates $\left(t,r,\theta,\phi\right)$  takes the form:
\begin{eqnarray}\label{doran}
d\tau^{2} & = & -\left(1-\frac{2Mr}{\rho^{2}}\right)dt^{2} + \frac{{\rho}^{2}}{r^{2}+a^{2}} dr^{2}- 2 \sqrt{\frac{2Mr}{r^{2}+a^{2}}} dt\: dr-4 a M r \frac{\sin^{2}{\theta}}{\rho^{2}} dt\: d\phi \\ 
& + & 2 a \sqrt{\frac{2Mr}{r^{2}+a^{2}}} \sin^{2}{\theta} d\phi dr 
+ \rho^{2} d\theta^{2}+\left[\left(r^{2}+a^{2}\right)+2Mra^{2}\frac{\sin^{2}{\theta}}{\rho^{2}}\right] {\sin{\theta}}^{2} {d\phi}^{2},\nonumber
\end{eqnarray}
where,
\begin{equation}
\rho^{2} = r^{2}+a^{2}{\cos^{2}{\theta}}.
\end{equation}
The constant $M$ represents the mass of the black hole and $a$ its angular momentum. In the limit $a \rightarrow 0$, Eq. (\ref{doran}) reduces to the Schwarzschild geometry in Gullstrand-Painlev\'e coordinates \cite{Gulls}\cite{Painle}.

Clifton et al.'s proposal is frame-dependent. We make the simplest choice for a Kerr spacetime (see below). Specifically, we adopt the following unit vectors $u^{a}$ and $z_{a}$:
\begin{eqnarray}
u^{a} & = & \left(0,\sqrt{\frac{r^2+a^2}{r^2+u^2}},0,0\right),\\
z_{a} & = &\left(1,0,0,0\right),\label{za}
\end{eqnarray}
where we have made the substitution:
\begin{equation}
u = a \cos{\theta}.
\end{equation}
The contravariant form of (\ref{za}) is
\begin{equation}
z^{a}  = \left(-1,-\frac{\sqrt{2Mr} \: \sqrt{r^2+a^2}}{r^2+u^2},0,0\right).
\end{equation}
This vector is chosen to be orthogonal to the hypersurfaces of constant time $t$\footnote{A vector field $\xi^{a}$is hypersurface orthogonal if and only if $\xi_{[a}\nabla_{b}{\xi_{c}}_{]} =0$ (see for instance \cite{Wald}). It can be shown that the vector $z_{a}$ given by Eq. (\ref{za}) satisfies $z_{[a}\nabla_{b}{z_{c}}_{]} =0$.}. It should be noticed that because of the four-fold coordinate degrees of freedom inherent to General Relativity, there is not a unique foliation of spacetime into a family of nonintersecting spacelike 3-surfaces $\Sigma$. For the Kerr spacetime metric given by Eq. (\ref{doran}), we have checked that the unit vectors $u^{a}$ and $z_{a}$ satisfy all conditions for the calculation of the gravitational entropy as stated by Clifton et al. \cite{Clifton}.

In the limit $a \rightarrow 0$, $u^{a}$ and $z_{a}$ do not coincide with the corresponding vectors given by Clifton et al. in Schwarzschild spacetime. As shown by Garat and Price \cite{Garat}, there is no spatial slicing of the Kerr spacetime that is both conformally flat and goes smoothly to a slice of constant Schwarzschild time as $a \rightarrow 0$. Hence, we do not neccesarily have to recover in the Schwarzschild limit the expressions for the gravitational energy density, gravitational temperature, and gravitational entropy obtained in \cite{Clifton}.

 
We now proceed to the calculation of the gravitational energy density and temperature according to Eqs. (\ref{dens-entropy}) and (\ref{T-grav}), respectively.

The gravitational energy density takes the following expression: 
\begin{equation}\label{dens-calc}
\rho_{\rm grav} = \frac{\alpha \ \sqrt{{\cal W}}}{2^{3/2}3^{1/2} \pi}.
\end{equation}
Our choice for the timelike unit vector $u^{a}$ leads to a simple form for the super-energy density ${\cal W}$:
\begin{equation}
{\cal W} = \frac{3\: M^{2}\left(3r^{2}-u^{2}\right)^{2}}{2\left(r^2+u^2\right)^{5}}.
\end{equation}

\begin{figure}[!htbp]
\begin{minipage}[b]{0.45\linewidth}
\centering
\includegraphics[width=7cm]{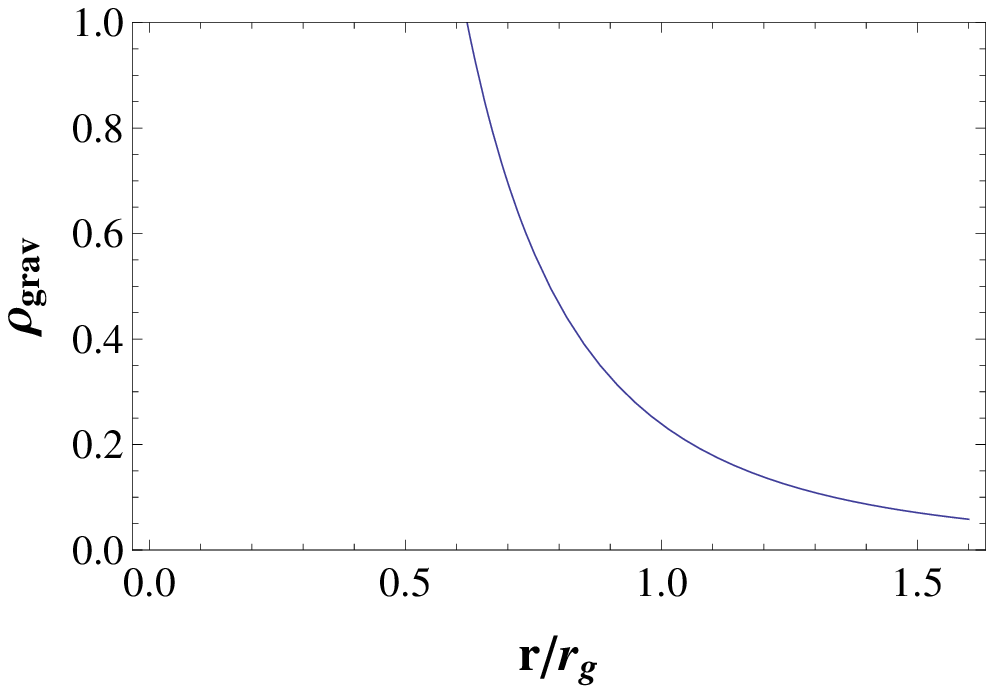}
\caption{Plot of $\rho_{\rm grav}$ as a function of the radial coordinate for $a = 0.8$ and $\theta = \pi /2$.}
\label{fig1-ref}
\end{minipage}
\hspace{0.45cm}
\begin{minipage}[b]{0.45\linewidth}
\centering
\includegraphics[width=7cm]{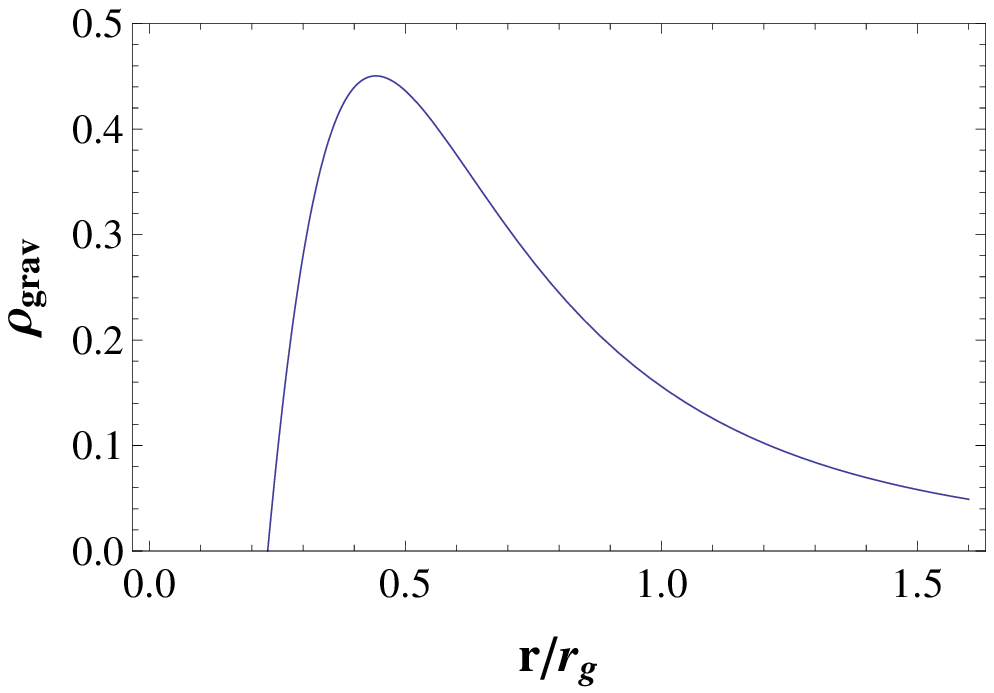}
\caption{Plot of $\rho_{\rm grav}$ as a function of the radial coordinate for $a = 0.8$ and $\theta = \pi /4$.}
\label{fig2-ref}
\end{minipage}
\end{figure}

We show in Figures \ref{fig1-ref} and \ref{fig2-ref} plots of $\rho_{\rm grav}$ as a function of the radial coordinate for $\theta = \pi/2$ and $\theta= \pi/4$, respectively. As expected, the gravitational energy density diverges towards the ring singularity ($r=0$, $\theta = \pi/2$). Outside the equatorial plane, $\rho_{\rm grav}$ is positive for $\left(3r^{2}-u^{2}\right)>0$, and has a maximum at $r = 11/9\; a \;{\cos^{2}{\theta}}$.

Then, the gravitational energy density $\rho_{\rm grav}$ is
\begin{equation}\label{dens-grav}
\rho_{\rm grav} = \frac{\alpha}{4 \pi}\frac{M\left(3r^{2}-u^{2}\right)}{\left(r^2+u^2\right)^{5/2}}.
\end{equation}

We now calculate the temperature of the gravitational field according to Eq. (\ref{T-grav}). The result is
\begin{equation}\label{T-calc}
T_{\rm grav} = \frac{\left|2\sqrt{2} \; r^{3/2}\sqrt{r^{2}+u^{2}}\left(2r^{2}+u^{2}+a^{2}\right)-\sqrt{M}\left(r^{2}+a^{2}\right)\left(r^{2}-u^{2}\right)\right|}{2^{3/2} \;  3 \; \pi \left(r^{2}+u^{2}\right)^{2}\sqrt{r\left(r^{2}+a^{2}\right)}},
\end{equation}
where the absolute value brackets were added to avoid negative or complex values. 

\begin{figure}[!ht]
\begin{minipage}[b]{0.45\linewidth}
\centering
\includegraphics[width=7cm]{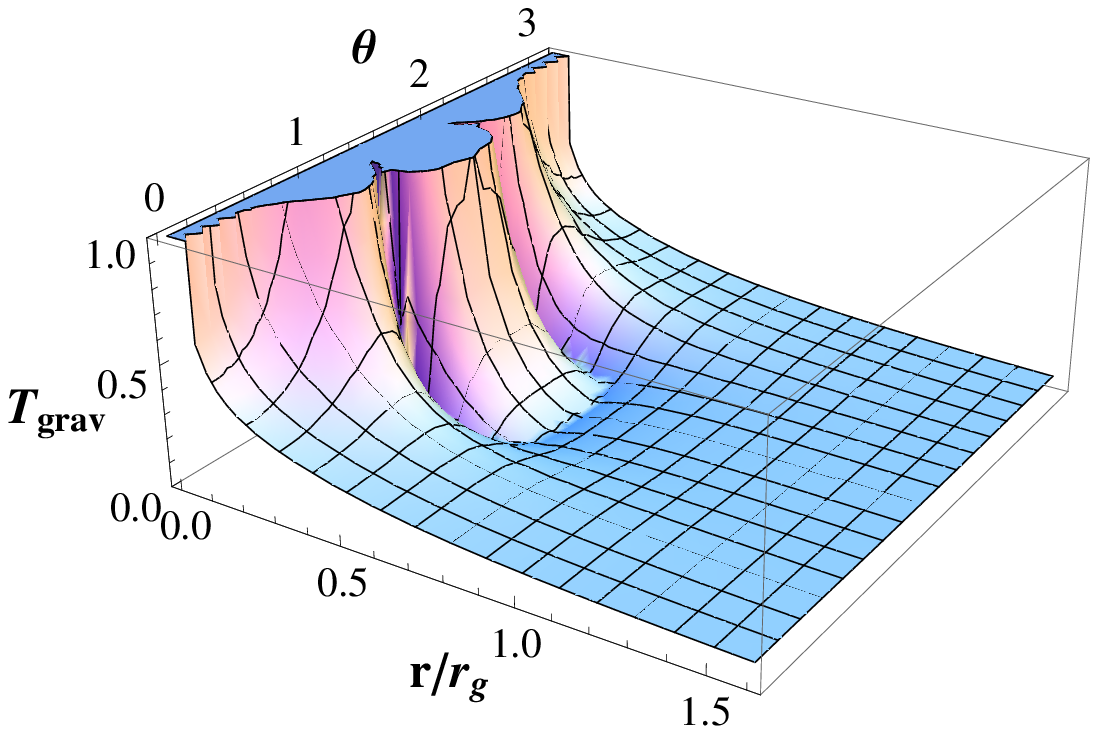}
\caption{Plot of $T_{\rm grav}$ as a function of the coordinates $r$ and $\theta$ for $a = 0.8$.}
\label{fig3-ref}
\end{minipage}
\hspace{0.45cm}
\begin{minipage}[b]{0.45\linewidth}
\centering
\includegraphics[width=7cm]{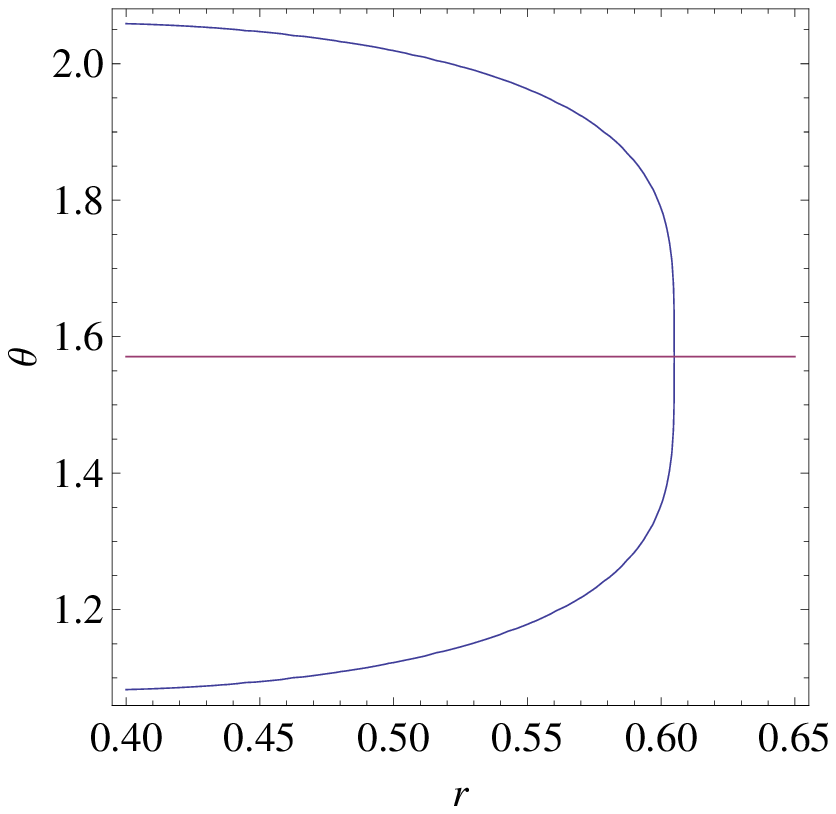}
\caption{Plot of the curve $f(r,\theta)=0$ for $a = 0.8$.}
\label{fig4-ref}
\end{minipage}
\end{figure}

We show in Figure \ref{fig3-ref} a 3-dimensional plot of $T_{\rm grav}$ as a function of the coordinates $r$ and $\theta$ for $a = 0.8$. We see that for $\theta \in \left[0,\pi\right]$, $T_{\rm grav}$ tends to infinity when $r$ goes to zero. For any other value of the radial coordinate the temperature of the gravitational field is well-defined.


As explained in \cite{Clifton}, a small change in the gravitational entropy density of a black hole occurs when a small amount of mass is added:
\begin{equation}\label{def-sgrav}
\delta s_{\rm grav} = \frac{\delta\left(\rho_{\rm grav} v\right)}{T_{\rm grav}}.
\end{equation}
In other words, a positive increment in the effective gravitational energy at a given $T_{\rm grav}$ causes a positive increment in the gravitational entropy. In the expression above the element of volume is
\begin{equation}
v = z^{a} \eta_{abcd} dx^{b} dx^{c} dx^{d},
\end{equation}
where $\eta_{abcd} = \eta_{\left[abcd\right]}$, $\eta_{0123} = \sqrt{\left|g_{ab}\right|}$. Given the Eqs. (\ref{doran}) and (\ref{za}), $v$ takes the form:
\begin{equation}
v = - \frac{\left(r^{2}+u^{2}\right)}{a} d\phi \ du \ dr.
\end{equation}

We now proceed to integrate Eq. (\ref{def-sgrav}) over the volume $V$ enclosed by the outer event horizon on a hypersurface of constant $t$, for a fixed value of $a$:
\begin{equation}\label{s-inte}
S_{\rm grav} = \int_{\rm V} \frac{\rho_{\rm grav} v}{T_{\rm grav}},
\end{equation} 
in order to test whether the Bel-Robinson proposal in the choosen frame reproduces the Bekenstein-Hawking entropy of a Kerr black hole. We notice, however, that independently of the coordinate choice, the region inside the inner horizon is not time-orientable since the region is chronology-violating (e.g. \cite{Visser}). The contribution to the gravitational entropy should come from the region between the inner and outer horizons.

Integral (\ref{s-inte}) can explicitly be written as:
\begin{eqnarray}
S_{\rm grav} & = &  3 \ 2^{1/2}\alpha \; M  {\int_{r_{-}}^{r_{+}}} {\int_{0}^{\pi}} {\int_{0}^{2 \pi}} s(r,\theta) \; d\phi \ d\theta \ dr, \nonumber \\
& = &  3 \ 2^{3/2} \pi \alpha \; M  {\int_{r_{-}}^{r_{+}}} {\int_{0}^{\pi}} \frac{\left(3r^{2}-u^{2}\right) \sqrt{r \left(r^{2}+a^{2}\right)\left(r^{2}+u^{2}\right)} \; \sin{\theta} \ d\theta \ dr}{\left|f(r,\theta)\right|}, \label{s-forma}
\end{eqnarray}
where in the latter expression we have integrated over the azimuthal coordinate $\phi$. 
The function $f(r,\theta)$ is defined as:
\begin{equation}
f(r,\theta) \equiv 2\sqrt{2} \; r^{3/2}\sqrt{r^{2}+u^{2}}\left(2r^{2}+u^{2}+a^{2}\right)-\sqrt{M}\left(r^{2}+a^{2}\right)\left(r^{2}-u^{2}\right).
\end{equation}
The domain of integration of Eq. (\ref{s-forma}) is:
\begin{equation}
T = \left\{\left(r,\theta\right) \in \Re^{2} / \  r_{-} \leq r \leq r_{+} \ \wedge \ 0 \leq \theta \leq \pi \right\}.
\end{equation}
We divide $T$ into two subregions denoted $D$ and $G$ respectively, such that
\begin{equation}
T = D \cup G,
\end{equation}
\begin{equation}
D = \left\{\left(r,\theta\right) \in \Re^{2} / \  r_{-} \leq r \leq r_{*} \ \wedge \ \theta_{*1} \leq \theta \leq \theta_{*2} \right\},
\end{equation}
where $r_{*}$, $\theta_{*1}$, and  $\theta_{*2}$ are the solutions of the following equations:
\begin{eqnarray}
f(r_{*},\pi/2) & = & 0,\\
f(r_{-},\theta_{*}) & = & 0,
\end{eqnarray}
and
\begin{equation}
G = G_{1} \cup G_{2} \cup G_{3},
\end{equation}
\begin{eqnarray}
G_{1} &  = & \left\{\left(r,\theta\right) \in \Re^{2} / \  r_{-} < r \leq r_{*} \ \wedge \ 0 \leq \theta \leq \theta_{*1} \right\},\\
G_{2} &  = & \left\{\left(r,\theta\right) \in \Re^{2} / \  r_{-} < r \leq r_{*} \ \wedge \ \theta_{*2} \leq \theta \leq \pi  \right\},\\
G_{3} &  = & \left\{\left(r,\theta\right) \in \Re^{2} / \  r_{*} < r \leq r_{+} \ \wedge \ 0 \leq \theta \leq \pi \right\}.
\end{eqnarray}
Given the definitions above, Eq. (\ref{s-forma}) can be written as:
\begin{equation}
S_{\rm grav} = {S}^{\rm D}_{\rm grav} + {S}^{\rm G}_{\rm grav},
\end{equation}
where,
\begin{eqnarray}
{S}^{\rm D}_{\rm grav} & = & 3 \ 2^{3/2} \pi \alpha \; M  {\int \int}_{D} \frac{\left(3r^{2}-u^{2}\right) \sqrt{r \left(r^{2}+a^{2}\right)\left(r^{2}+u^{2}\right)} \; \sin{\theta} \ d\theta \ dr}{\left|f(r,\theta)\right|},\label{s-inte-D}\\
{S}^{\rm G}_{\rm grav} & = & 3 \ 2^{3/2} \pi \alpha \; M  {\int \int}_{G} \frac{\left(3r^{2}-u^{2}\right) \sqrt{r \left(r^{2}+a^{2}\right)\left(r^{2}+u^{2}\right)} \; \sin{\theta} \ d\theta \ dr}{\left|f(r,\theta)\right|}.\label{s-inte-G}
\end{eqnarray}
If $0 < a < 0.94$, the integral given by Eq. (\ref{s-inte-D}) is an improper divergent integral\footnote{For a formal definition of improper integral see the Appendix and Ref. \cite{Courant}.}; in particular it tends to infinity for those values of $r$ and $\theta$ such that $f(r,\theta)= 0$. We show in Figure \ref{fig4-ref} a plot of the curve $f(r,\theta) = 0$ for $a=0.8$. Contrary, integral (\ref{s-inte-G}) is well defined for $a \in (0,1)$. If $0.94 \leq a < 1$, ${S}^{\rm D}_{\rm grav}$ is not an improper divergent integral, and can be integrated numerically as we will show below.

\begin{figure}
\centering
\includegraphics[width=8cm]{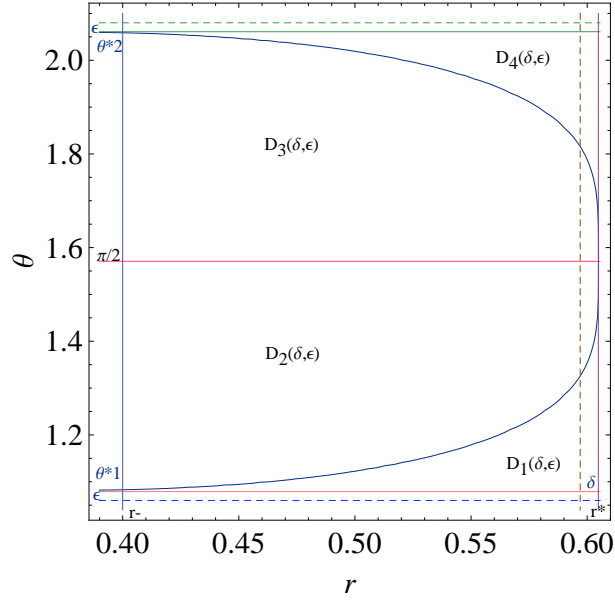}
\caption{Plot of the integration domain $D_{\left(\delta,\epsilon\right)}$.}
\label{graf-domain}
\end{figure}

In order to prove that integral (\ref{s-inte-D}) is divergent for $0 < a < 0.94$, we define four closed subregions: 
\begin{equation}
D_{\left(\delta,\epsilon\right)} = D_{1 \left(\delta,\epsilon\right)} \cup D_{2 \left(\delta,\epsilon\right)} \cup D_{3 \left(\delta,\epsilon\right)} \cup D_{4 \left(\delta,\epsilon\right)},
\end{equation}
\begin{eqnarray}
D_{1 \left(\delta,\epsilon\right)} & = & \left\{\left(r,\theta\right) \in \Re^{2} / \  r_{-} \leq r \leq r_{*}-\delta  \ \wedge \ \theta_{*1}-\epsilon \leq \theta \leq \pi/2  \ \wedge f(r,\theta)> 0 \right\},\\
D_{2 \left(\delta,\epsilon\right)} & = & \left\{\left(r,\theta\right) \in \Re^{2} / \  r_{-} \leq r \leq r_{*}-\delta  \ \wedge \ \theta_{*1}-\epsilon \leq \theta \leq \pi/2  \ \wedge f(r,\theta)< 0 \right\},\\
D_{3 \left(\delta,\epsilon\right)} & = & \left\{\left(r,\theta\right) \in \Re^{2} / \  r_{-} \leq r \leq r_{*}-\delta  \ \wedge \ \pi/2 \leq \theta \leq \theta_{*2}+\epsilon  \ \wedge f(r,\theta) < 0 \right\},\\
D_{4 \left(\delta,\epsilon\right)} & = & \left\{\left(r,\theta\right) \in \Re^{2} / \  r_{-} \leq r \leq r_{*}-\delta  \ \wedge \ \pi/2 \leq \theta \leq \theta_{*2}+\epsilon  \ \wedge f(r,\theta) > 0 \right\},\\
\end{eqnarray}
where $\epsilon,\delta > 0$, and
\begin{equation}
\lim_{\left(\delta,\epsilon\right)\rightarrow 0} D_{\left(\delta,\epsilon\right)} \rightarrow D.
\end{equation}
A plot of the domain of integration $D_{\left(\delta,\epsilon\right)}$ is shown in Figure \ref{graf-domain}. In each $D_{\left(\delta,\epsilon\right)}$ the function $s(r,\theta)$ is well defined and continuous:
\begin{displaymath}
s\left(r,\theta\right) = \left\{\begin{array}{ll}
s_{1}\left(r,\theta\right) = \frac{\left(3r^{2}-u^{2}\right) \sqrt{r \left(r^{2}+a^{2}\right)\left(r^{2}+u^{2}\right)}\; \sin{\theta}}{2\sqrt{2} \; r^{3/2}\sqrt{r^{2}+u^{2}}\left(2r^{2}+u^{2}+a^{2}\right)-\sqrt{M}\left(r^{2}+a^{2}\right)\left(r^{2}-u^{2}\right)},  & \textrm{for $(r,\theta) \in D_{1 \left(\delta,\epsilon\right)}$ }\\ \\
s_{2}\left(r,\theta\right) =  - \frac{\left(3r^{2}-u^{2}\right) \sqrt{r \left(r^{2}+a^{2}\right)\left(r^{2}+u^{2}\right)}\; \sin{\theta}}{2\sqrt{2} \; r^{3/2}\sqrt{r^{2}+u^{2}}\left(2r^{2}+u^{2}+a^{2}\right)-\sqrt{M}\left(r^{2}+a^{2}\right)\left(r^{2}-u^{2}\right)}, & \textrm{for $(r,\theta) \in D_{2 \left(\delta,\epsilon\right)}$ }\\ \\
s_{3}\left(r,\theta\right) = \frac{\left(3r^{2}-u^{2}\right) \sqrt{r \left(r^{2}+a^{2}\right)\left(r^{2}+u^{2}\right)} \; \sin{\theta}}{2\sqrt{2} \; r^{3/2}\sqrt{r^{2}+u^{2}}\left(2r^{2}+u^{2}+a^{2}\right)-\sqrt{M}\left(r^{2}+a^{2}\right)\left(r^{2}-u^{2}\right)}, & \textrm{for $(r,\theta) \in D_{3 \left(\delta,\epsilon\right)}$ }\\ \\
s_{4}\left(r,\theta\right) = -\frac{\left(3r^{2}-u^{2}\right) \sqrt{r \left(r^{2}+a^{2}\right)\left(r^{2}+u^{2}\right)} \; \sin{\theta}}{2\sqrt{2} \; r^{3/2}\sqrt{r^{2}+u^{2}}\left(2r^{2}+u^{2}+a^{2}\right)-\sqrt{M}\left(r^{2}+a^{2}\right)\left(r^{2}-u^{2}\right)},  & \textrm{for $(r,\theta) \in D_{4\left(\delta,\epsilon\right)}$ }\\ \\
\end{array} \right.
\end{displaymath}
The integral of the function $s(r,\theta)$ over the domain $D_{\left(\delta,\epsilon\right)}$  takes the form:
\begin{eqnarray}
{\int \int}_{D_{\left(\delta,\epsilon\right)} } s\left(r,\theta\right) d\theta \ dr  & = & {\int \int}_{D_{1 \left(\delta,\epsilon\right)}} s_{1}\left(r,\theta\right) d\theta \ dr  + {\int \int}_{D_{2 \left(\delta,\epsilon\right)}}s_{2}\left(r,\theta\right) d\theta \ dr  + {\int \int}_{D_{3 \left(\delta,\epsilon\right)}} s_{3}\left(r,\theta\right) d\theta \ dr  \nonumber \\
& + & {\int \int}_{D_{4 \left(\delta,\epsilon\right)}} s_{4}\left(r,\theta\right) d\theta \ dr. \nonumber 
\end{eqnarray}

The function $f(r,\theta)$ can be well approximated by a power series expansion about the point $r=r_{*}$ to order $\left(r-r_{*}\right)^{4}$ in the region $D_{\left(\delta,\epsilon\right)}$:
\begin{eqnarray}
f(r,\theta) & \cong f_{4}& \equiv a_{0}\left(a,r_{*},\theta\right)+ a_{1}\left(a,r_{*},\theta\right) \left(r_{*}-r\right)+ a_{2}\left(a,r_{*},\theta\right)\left(r_{*}-r\right)^{2}\\ \nonumber
& + & a_{3}\left(a,r_{*},\theta\right)\left(r_{*}-r\right)^{3} + a_{4}\left(a,r_{*},\theta\right)\left(r_{*}-r\right)^{4},
\end{eqnarray}
where the coefficients $a_{j}\left(a,r_{*},\theta\right)$, $j = 0,..,4$ are continuous functions of $a$, $r_{*}$, and $\theta$ in each $D_{i\left(\delta,\epsilon\right)}$, $i =1,..,4$.

Since $D_{\left(\delta,\epsilon\right)}$ is a closed and bounded set and  $f_{4}$ is continuous in $D_{\left(\delta,\epsilon\right)}$, then there exists points in 
$D_{\left(\delta,\epsilon\right)}$ where each coefficient $a_{j}$,  $j = 0,..,4$ has its maxima and also points where each $a_{j}$ has minima \cite{Courant}. We make use of this property to find in each domain $D_{i\left(\delta,\epsilon\right)}$, $i =1,..,4$ a constant $\tilde{a}^{i}_{j} \geq 0 $ such that
\begin{equation}
\sum_{j = 0}^{4} a_{j} \left(r_{*}-r\right)^{j} \leq \sum_{j = 0}^{4} \tilde{a}^{i}_{j} \left(r_{*}-r\right)^{j} \equiv g_{i}\left(r\right) \ \ \ \ \ \ \ \ \ \forall \ i=1,..,4.
\end{equation}
In the latter equation $g_{i}\left(r\right)$ represents a polynomial of degree four. Since $r<1$ in each $D_{i\left(\delta,\epsilon\right)}$, $i =1,..,4$, we can always find a linear function of the form
\begin{equation}
h_{i}(r) \equiv \gamma_{i} \left(r_{*}-r\right) \ \ \ \ \ \gamma_{i} \in \Re^{+},
\end{equation}
such that
\begin{equation}
g_{i}\left(r\right) \leq h_{i}(r) \ \ \ \ \ \ \ \ \ \forall \ i=1,..,4.
\end{equation} 
Then,
\begin{equation}
0 \leq \frac{1}{h_{i}(r)} \leq \frac{1}{g_{i}\left(r\right)} \leq \frac{\tilde{\mathbb{C}_{i}}}{f_{4}} \leq s_{i}\left(r,\theta\right) \ \ \ \forall \ i=1,..,4\ \ \ \ \ \tilde{\mathbb{C}_{i}} \in \Re^{+},
\end{equation}
and (see \cite{Courant} p. 383):
\begin{equation}\label{des-int}
{\int \int}_{D_{i\left(\delta,\epsilon\right)}} \frac{1}{h_{i}(r)} d\theta\ dr \leq {\int \int}_{D_{i\left(\delta,\epsilon\right)}}  \frac{\tilde{\mathbb{C}_{i}}}{f_{4}} d\theta\ dr \leq {\int \int}_{D_{i\left(\delta,\epsilon\right)}} s_{i}\left(r,\theta\right) d\theta\ dr.
\end{equation}

It is proved in the Appendix that the integral
\begin{equation}
I \equiv {\int \int}_{D} \frac{1}{h_{i}(r)} d\theta\ dr,\ \ \ \forall \ i=1,..,4,
\end{equation} 
is divergent. Hence, we can state that since in each $D_{i\left(\delta,\epsilon\right)}$, $i =1,..,4$ the inequality given by (\ref{des-int}) holds, then:
\begin{equation}
\sum_{i = 1}^{4} {\int \int}_{D_{i\left(\delta,\epsilon\right)}}  \frac{1}{h_{i}(r)} d\theta\ dr \leq \sum_{i = 1}^{4} {\int \int}_{D_{i\left(\delta,\epsilon\right)}} s_{i}\left(r,\theta\right) d\theta\ dr = {\int \int}_{D_{\left(\delta,\epsilon\right)}} s\left(r,\theta\right) d\theta\ dr.
\end{equation}
Application of the comparison criterion for improper integrals \cite{Predoi} leads us to conclude that ${S}^{\rm D}_{\rm grav}$ is divergent.

\begin{figure}
\centering
\includegraphics[width=12cm]{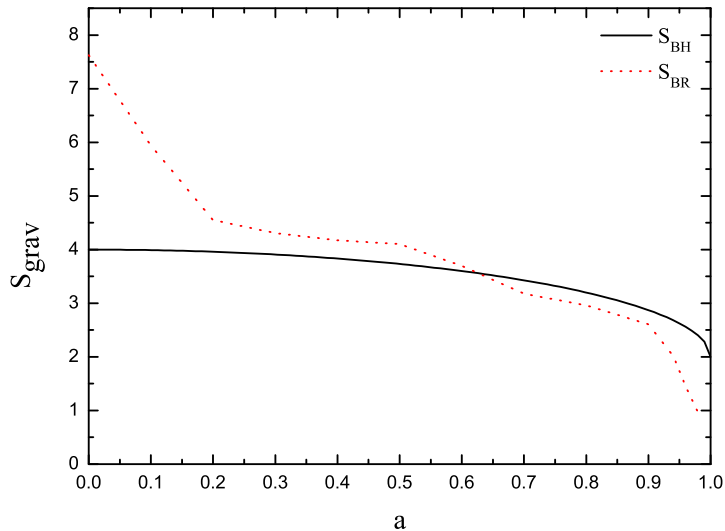}
\caption{Plot of  ${S_{\rm BR}}$ and ${S_{\rm BH}}$ as a function of the angular momentum $a$.}
\label{compa-s}
\end{figure}

We now integrate numerically ${S}^{\rm G}_{\rm grav}$ (see Eq. \ref{s-inte-G}) for $0 < a < 0.94$, and Eq. (\ref{s-forma}) for $0.94 \leq a < 1$. The result is shown in Figure \ref{compa-s}. We plot the Bel-Robinson gravitational entropy, denoted ${S_{\rm BR}}$ and also the Bekenstein-Hawking entropy ${S_{\rm BH}}$ as a function of the angular momentum of the hole. It is clear that ${S_{\rm BR}}$ does not reproduce the Bekenstein-Hawking entropy of a black hole. We conclude that even in the small range of the angular momentum where the entropy is well defined, it is not a good approximation to the Bekenstein-Hawking entropy, at least for the current coordinate choice.

Let us now briefly compare the situation with the Rudjord-Gr$\varnothing$n-Hervik picture.

The calculation of the Weyl-Kretschmann gravitational entropy for Kerr spacetimes was done by Romero, Thomas, and P\'erez \cite{Romero}. This estimator is defined (see Subsection \ref{sub:WK}) in such a way that it necessarily matches the Bekenstein-Hawking entropy in the event horizon of the black hole. This is not the case for the Bel-Robinson estimator.

\begin{figure}[h!]
\begin{center}
\includegraphics[width=10cm]{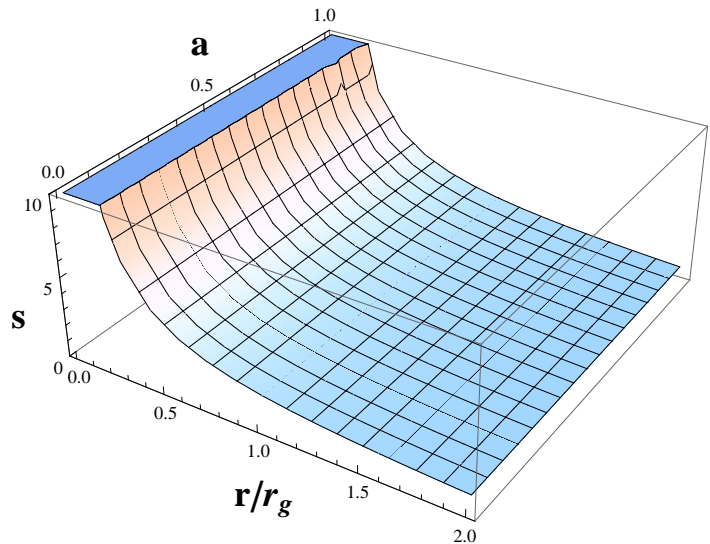}
\end{center}
\caption{Weyl-Kretschmann gravitational entropy density of a Kerr black hole as a function of the radial coordinate and the angular momentum for $\theta = \pi/2$.}
\label{densitykerr1}
\end{figure}

For the Weyl-Kretschmann proposal, the gravitational entropy density is
\begin{equation}\label{WK-entro}
s = k \left|\frac{4r}{a^{2}+2r^{2}+a^{2}\cos{2\theta}}\right|.
\end{equation}
A plot of Eq. (\ref{WK-entro}) is shown in Figure \ref{densitykerr1}. We see that $s$ is everywhere well-defined except where the spacetime is singular ($r = 0$, $\theta = \pi/2$). At large distances, $r \rightarrow \infty$, $s$ tends to zero, as expected, contrary to what happens with Bel-Robinson estimator.

\section{Closing remarks}

We have computed the gravitational energy density, temperature, and gravitational entropy of a Kerr black hole according to the Bel-Robinson estimator. Our calculations are based on a specific choice of a timelike unit vector $u_{a}$ and a spacelike unit vector $z^{a}$ that determine a Weyl principal tetrad. The choice of such vectors, however, is not unique. Consequently, $\rho_{\rm grav}$, $T_{\rm grav}$ are frame dependent quantities. We proved that, with the simplest coordinate choice, the gravitational entropy does not reproduce the Bekenstein-Hawking entropy of a Kerr black hole. We do not discard that for a different choice of vectors $u_{a}$ and $z^{a}$, the Bel-Robinson proposal may coincide with Bekenstein-Hawking result. However, the fact that the innermost region of the Kerr spacetime is not folliable and time-orientable suggests that our result might be general.

The Weyl-Kretschmann estimator works according expectations for Kerr and Kerr-Newman spacetimes \cite{Romero}. An advantage of this latter estimator with respect to Bel-Robinson's is that it is defined in such a way that is not dependent on a specific folliation of spacetime and can be applied to chronology-violating spacetimes such as those of Kerr and Kerr-Newman, or even wormhole spacetimes \cite{Romero}.

It is clear that more research is needed in this field if we want to have at our disposal a reliable classical estimator of the gravitational entropy. The first requirement that needs to be fulfilled is that the estimator be well-behaved in all types of horizons where quantum field calculations can be used as an independent probe of the entropy. Only when a complete match be obtained, the classical estimators can be used to evaluate other families of spacetimes with some confidence.

\section*{Acknowledgements}

This work is supported by PICT 2012-00878, Pr\'estamo BID (ANPCyT). We thank Santiago E. Perez Bergliaffa for useful advice.


\begin{thebibliography}{}
%


\bibitem{Hawking1}
Hawking, S. W.: Black hole explosions? Nature 248, 30-31 (1974)

\bibitem{Hawking2}
Hawking, S. W.: Particle creation by black holes. Commun. Math. Phys. 43, 199-220 (1975)

\bibitem{Bekenstein1}
Bekenstein, J. D.: Black holes and the second law. Lett. Nuovo Cim. 4, 737-740 (1972)

\bibitem{Bekenstein2}
Bekenstein, J. D.: Black holes and entropy. Phys. Rev. D 7, 2333-2346 (1973)

\bibitem{Bekenstein3}
Bekenstein, J. D.: Generalized second law of thermodynamics in black holes, Phys. Rev. D 9, 3292-3300 (1974)

\bibitem{Bardeen}
Bardeen, J. M., Carter, B., Hawking, S. W.: The four laws of black hole mechanics. Commun. Math. Phys. 31, 161-170 (1973)

\bibitem{Penrose}
Penrose, R.: General Relativity, an Einstein Centenary Survey. In: Hawking, S.W., Israel, W. (eds.) Singularity
and Time-Asymmetry, pp. 581-638. Cambridge Univ. Press, Cambridge (1979)

\bibitem{Gron}
Rudjord, $\varnothing$, Gr$\varnothing$n, $\varnothing$., Hervik, S.: The Weyl curvature conjeture and black hole entropy. Phys. Scr. 77 Issue 5, 055901, 1-7 (2008)


\bibitem{Clifton}
Clifton, T., Ellis, G.F.R., Tavakol, R.: A gravitational entropy proposal. Class. Quant. Grav., 30, 125009 (2013)

\bibitem{Bel1}
Bel, L.: Sur la radiation gravitationnelle. C. R. Acad. Sci. 247, 1094-1096 (1958) 

\bibitem{Bel2}
Bel, L.: Introduction d'un tenseur du quatrieme order. C. R. Acad. Sci. 248, 1297 (1959) 

\bibitem{Robinson1}
Robinson, I., unpublished lectures at King College, London (1958)

\bibitem{Robinson2}
Robinson, I.: On the Bel-Robinson tensor. Class. Quant. Grav. 14, 4331 (1997)

\bibitem{Breton}
Breton, N., Feinstein, A., Iba\~{n}ez, J.: The Bel-Robinson tensor for the collision of gravitational plane waves. Gen. Rel. Grav. 25, 267-273 (1993)

\bibitem{Krish}
Krishnasamy, I.: Quasilocal energy and the Bel-Robinson tensor. Gen. Rel. Grav. 17, 621-627 (1985)

\bibitem{Bonilla}
Bonilla, M. A. G., Senovilla, J. M. M.: Some properties of the Bel and Bel-Robinson tensors. Gen. Rel. Grav. 29, 91-116 (1997)

\bibitem{Unruh}
Unruh, W. G.: Notes on black-hole evaporation. Phys. Rev. D 14, 870-892 (1976)

\bibitem{deSitter}
Gibbons, G. W., Hawking, S. W.: Cosmological event horizons, thermodynamics, and particle creation. Phys. Rev. D 15, 2738-2751 (1977)

\bibitem{Doran}
Doran, C.: A new form of the Kerr solution. Phys. Rev. D 61, 067503 (2000)


\bibitem{Gulls}
Gullstrand, A.: Allgemeine l$\ddot{o}$sung des statichen eink$\ddot{o}$rperproblems in der Einsteinschen gravitationstheorie. Arkiv. Mat. Astron. Fys. 16, 1–15 (1922)

\bibitem{Painle}
Painlev\'e, P.: La m\'ecanique classique et la th\'eorie de la relativit\'e, C. R. Acad. Sci. (Paris) 173, 677–680 (1921)

\bibitem{Wald}
Wald, R. M.: General Relativity. The University of Chicago Press, Chicago (1984)

\bibitem{Garat}
Garat, A., Price, R. H.: Nonexistence of conformally flat slices of the Kerr spacetime. Phys. Rev. D 61, id. 124011 (2000)

\bibitem{Visser}
Visser, M.: Lorentzian wormholes: from Einstein to Hawking. AIP Series in Computational and Applied Mathematical Physics. Springer-Verlag, New York (1996)


\bibitem{Courant}
Courant, R.: Differential and integral calculus Vol II. Blackie \& Son, Great Britain (1961)

\bibitem{Predoi}
Predoi, M., B$\breve{a}$lan, T.: Mathematical analysis Vol II. Integral Calculus. Editura Universitaria, Craiova (2005)

\bibitem{Romero}
Romero, G. E., Thomas, R., P\'erez D.: Gravitational entropy of black holes and wormholes. Int. J. Theor. Phys. 51, 925–942 (2012)


\end{thebibliography}

\appendix
\section*{\textbf{Appendix}}
\section*{Definition of improper integral}

Let $\Omega \subseteq \Re^{p}$ be a non-compact domain, and let $f: \Omega \rightarrow \Re$ be integrable on each measurable compact domain $D \subset \Omega$. We say that $f$ is \textsl{improperly integrable} on $\Omega$ iff for every increasing sequence of measurable compact domains, $\left(D_{\rm n}\right)_{n \in \mathbb{N}}$, which is exhausting\footnote{Let $\Omega \subseteq \Re^{p}$ be a non-compact domain for which each bounded part of the frontier is negligible. We say that the sequence $\left(D_{\rm n}\right)_{n \in \mathbb{N}}$ of measurable compact domains is \textsl{exhausting} $\Omega$ iff for any compact $K \subset \Omega$ there exists $n_{0} \in \mathbb{N}$ such that $K \subset D_{\rm n}$ for all $n \geq n_{0}$.} $\Omega$, the secuence:
\begin{displaymath}
\sum_{\rm n} = \left(\int_{D_{\rm n}} f \ d\mu\right)_{\rm n \in \mathbb{N}}\nonumber
\end{displaymath}
is convergent. In such a case we note:
\begin{displaymath}
\lim_{{\rm n} \rightarrow \infty}  \int_{D_{\rm n}} f \ d\mu = \int_{\Omega} f \ d\mu,\nonumber
\end{displaymath}
and we call it \textsl{improper integral of $f$ on $D$}. Alternatively we say that the integral of $f$ on $\Omega$ is \textsl{convergent} \cite{Predoi}.

\section*{Proof of the divergence of the improper integral:}

\begin{equation}\label{diver}
I = {\int \int}_{D} \frac{1}{h_{i}(r)} d\theta\; dr = {\int \int}_{D} \frac{1}{\gamma_{i} \left(r_{*}-r\right)} \ \ \ \ \ \ \ \ \gamma_{i} \in \Re^{+},
\end{equation} 
where
\begin{equation}
D = \left\{\left(r,\theta\right) \in \Re^{2} / \  r_{-} \leq r \leq r_{*} \ \wedge \ \theta_{*1} \leq \theta \leq \theta_{*2} \right\},
\end{equation}

Since $h_{i}(r)$ does not depend on $\theta$, we can simply integrate (\ref{diver}) on this coordinate:
\begin{equation}
\int_{R} \frac{\theta_{*2}-\theta_{*1}}{\gamma_{i} \left(r_{*}-r\right)} dr,
\end{equation}
where,
\begin{equation}
R = \left\{ r \in \Re /  r_{-} \leq r < r_{*}\right\}.
\end{equation}
The function $1/\gamma_{i} \left(r_{*}-r\right)$ diverges at $r = r*$. We define a subsequence of closed subregions $R_{\rm n}$ where the latter integral is well-defined:
\begin{equation}
R_{n} = \left\{ r \in \Re /  r_{-} \leq r < r_{*}-\left(1/n\right)\right\},
\end{equation}
with $n \in {\mathbb N}$, such that
\begin{equation}
\lim_{n \rightarrow \infty} R_{n} \rightarrow R.
\end{equation}

We integrate the function $1 = 1/h_{i}(r)$ over the domain $R_{n}$ as follows:
\begin{equation}
{\int}_{R_{n}} \frac{dr}{h_{i}(r)} = {\int}_{R_{n}} \frac{dr}{\gamma_{i} \left(r_{*}-r\right)}  = \frac{-1}{\gamma_{i}} \ln{\left(r_{*}-r\right)} \ \vert_{r_{-}}^{r_{*}-\left(1/n\right)} = \frac{-1}{\gamma_{i}} \left[\ln\left(r_{*}-r_{*}+(1/n)\right)-\ln\left(r_{*}-r_{-}\right)\right].
\end{equation}
The limit of the latter equation when $n \rightarrow \infty$ does not exist. Therefore, the integral given by Eq. (\ref{diver}) is divergent.


\end{document}